\documentclass[twocolumn]{aastex63}

\usepackage{gensymb}
\usepackage[figuresright]{rotating}

\graphicspath{{./}{figures/}}

\usepackage{hyperref}

\begin{document}

\title{Investigating the Sulfur Mystery in Protoplanetary Disks Through Chemical Modeling}


%
\correspondingauthor{Becky J. Williams}
\email{rjw9dmj@virginia.edu}

\author[0000-0002-1548-6811]{\textbf{Becky J. Williams}}
\affiliation{Department of Astronomy, University of Virginia, Charlottesville, VA 22904, USA}

\author[0000-0003-2076-8001]{\textbf{L. Ilsedore Cleeves}}
\affiliation{Department of Astronomy, University of Virginia, Charlottesville, VA 22904, USA}
\affiliation{Department of Chemistry, University of Virginia, Charlottesville, VA 22904, USA}

\author[0000-0002-0477-6047]{\textbf{Rachel E. Gross}}
\affiliation{Department of Chemistry, University of Virginia, Charlottesville, VA 22904, USA}

\author[0009-0008-7850-1485]{\textbf{Jackson Baker}}
\affiliation{Department of Astronomy, University of Virginia, Charlottesville, VA 22904, USA}



\begin{abstract}

Sulfur is a critical element to life on Earth, and with detections of sulfur-bearing molecules in exoplanets and comets, questions arise as to how sulfur is incorporated into planets in the first place. In order to understand sulfur's journey from molecular clouds to planets, we need to understand the molecular forms that sulfur takes in protoplanetary disks, where the rotational emission from sulfur-bearing molecules in the gas phase indicates a very low abundance. To address this question, we have updated the 2D time-dependent disk chemical modeling framework of \citet{Fogel2011} to incorporate several new sulfur species and hundreds of new sulfur reactions from the literature. Specifically, we investigate the main molecular forms that sulfur takes in a disk orbiting a solar mass young T Tauri star. We explore the effects of different volatile (reactive) sulfur abundances, C/O ratios, initial sulfur molecular forms, and cosmic-ray ionization rates. We find that a high C/O ratio can explain both the prevalence of CS observed in disks and the lack of SO detections, consistent with previous results. Additionally, initial sulfur form greatly affects the ice abundances in the lower layers of the disk, which has implications for comet formation and future observations with JWST.

\end{abstract}



\section{Introduction} \label{sec:intro}

One of the keys to understanding the emergence of life is determining the history of life's required elements. Sulfur is one such element, thought to have played a role in the origin of life on Earth. For example, sulfur-bearing molecules in early Earth's atmosphere may have provided protection against harmful solar radiation \citep{Kasting1989}, and evidence suggests that early microbial life metabolized sulfur \citep{Lake1988}. Sulfur-bearing molecules remain important to life today: some microorganisms use hydrogen sulfide (H$_2$S) for anoxygenic photosynthesis \citep{Kushkevych2021}, and the sulfur-containing amino acids methionine and cysteine are both important constituents of proteins \citep{Brosnan2006}. 

Clearly, the story of how life arose on Earth requires understanding how sulfur is incorporated into planets. This question is even more intriguing considering the recent detection of SO$_2$ in the atmosphere of exoplanet WASP-39b \citep{Rustamkulov2023} and the variety of sulfur molecules, including H$_2$S, SO$_2$, SO, OCS, and H$_2$CS, detected in the coma of comet 67P/Churyumov-Gerasimenko \citep{Calmonte2016, LeRoy2015}. To understand how sulfur was incorporated into these environments, we need to understand the molecular forms and distribution of sulfur in protoplanetary disks.

The solar sulfur abundance relative to hydrogen is $1.3\times10^{-5}$ \citep{Asplund2021}. In protoplanetary disks, the majority of sulfur is predicted to be in refractory form, such as dust. Using observations of stellar photospheres of young stars, \citet{Kama2019} measured that $\approx89\pm8\%$ of sulfur is in a nonobservable and nonreactive refractory form such as FeS.

Several gas-phase sulfur-bearing molecules have been detected in low abundance in disks. The most widely detected is CS, which has been observed in several disks, including IM Lup, GM Aur, AS 209, HD 163296, MWC 480, and TW Hya \citep{LeGal2021, Teague2018a}. Other gas-phase molecules that have been detected in disks include SO, H$_2$S, H$_2$CS, SO$_2$, and C$_2$S \citep{Fuente2010,RiviereMarichalar2022,LeGal2021,Booth2021,Phuong2021}.

\citet{Semenov2018} searched for CS, SO, SO$_2$, OCS, C$_2$S, H$_2$CS, and H$_2$S in DM Tau, and only detected CS, and tentatively SO$_2$. They suggested that a C/O ratio $\gtrsim 1$ is needed to explain the inferred high abundance of CS relative to other species, specifically SO and SO$_2$. This observation is likely linked to the depletion of oxygen observed in disks \citep[e.g.,][]{Dutrey1994,Ansdell2016,Long2017,Du2017}. Modeling by \citet{LeGal2021} and \citet{Keyte2023} has shown that the C/O ratio has a large effect on the resulting abundances of sulfur-bearing species, including the CS/SO ratio.

A number of sulfur detections are linked to unique morphologies and processes. CS, SO, and H$_2$S have all been detected in AB Auriga, a Herbig Ae star, whose disk has an asymmetric continuum. Abundances of these three molecules are on the order of $10^{-11}$ to $10^{-10}$ relative to hydrogen \citep{Fuente2010,RiviereMarichalar2022}. \citet{Law2023} detected SO and SiS in HD 169142, potentially tracing temperature variation and outflow due to a young planet. \citet{Booth2021} detected SO and SO$_2$ in Oph-IRS 48, a Herbig source that has a dust trap. Both SO and SO$_2$ emission spatially coincide with the location of the dust trap, while CO emission covers the full azimuth extent of the disk. \citet{Booth2021} calculated a detected S/H ratio of $4.6-10.0\times10^{-7}$, which accounts for $\approx15-100\%$ of the total volatile sulfur budget predicted by \citet{Kama2019}. They propose that the high abundances of SO and SO$_2$ are due to ice sublimation from the dust trap. In contrast to observations of other disks, they do not detect CS in Oph-IRS 48, and therefore calculate a very low CS/SO $< 0.01$.

\citet{Keyte2024} studied the disk HD 100546, using observations from the Atacama Large Millimeter/submillimeter Array (ALMA) and the Atacama Pathfinder EXperiment (APEX), as well as chemical modeling. They determined that the volatile S/H $\sim10^{-8}$, and predict that the main gas-phase carriers are OCS, H$_2$CS, and CS. They also predict a large OCS ice reservoir.

Although observations of disks have yielded several detections of sulfur-bearing molecules, the vast majority of sulfur remains undetected in protoplanetary disks, and several questions remain as to the evolution and abundance of sulfur throughout the protoplanetary disk phase. Chemical modeling is one way in which we can probe this hidden sulfur reservoir.

In recent years, efforts have been made to improve the chemical modeling of sulfur species, especially for regions of the interstellar medium. Much of this effort has been motivated by the Gas phase Elemental abundances in Molecular cloudS (GEMS) program, which aims to determine the elemental abundances of several elements, including sulfur \citep{Fuente2023}, in the interstellar medium. \citet{Fuente2016,Fuente2019} and \citet{Rocha2023} computed new rate constants for five reactions involving SO and CS. \citet{Vidal2017} presented an updated sulfur network with values from the literature and used this updated network to model dark clouds. \citet{Vastel2018} used the \citet{Vidal2017} network to model a prestellar core, and \citet{Laas2019paper} incorporated much of the \citet{Vidal2017} network into their network, which they then used to model diffuse, translucent, and dense molecular clouds. More recently, \citet{Santos2024} studied the ice reactions that result from combining C$_2$H$_2$, H$_2$S, and atomic H, finding that the main sulfur-bearing product is CH$_3$CH$_2$SH.

Motivated by this growing body of observational and theoretical work on sulfur chemistry, we have updated the disk chemical model originally presented in \citet{Fogel2011} with this new information and present an exploration of how the forms that sulfur takes in protoplanetary disks change under different physical and chemical assumptions.

\section{Methods} \label{sec:methods}

\subsection{Chemical Model} \label{met1}

The model utilized in this paper was first presented in \citet{Fogel2011} and updated in \citet{Cleeves2018} and \citet{Anderson2021}. The model is 2D in radius and height and calculates the time-evolving abundances with no vertical or horizontal movement of material. This model is unique because it incorporates radiation fields in a wavelength-dependent manner and calculates the radiative transfer throughout the disk. Additionally, the model includes photodissociation and photoionization of molecules based on their wavelength-dependent cross sections where available \footnote{\url{https://home.strw.leidenuniv.nl/~ewine/photo/}}. The sulfur-bearing molecules that have cross sections included in the model are CS, SO, SO$_2$, CS$_2$, OCS, H$_2$S, D$_2$S, and HDS. We run the model up to 1 Myr, with 60 logarithmically spaced time steps, and we analyze the output at 1 Myr.

\subsection{Reaction Network} \label{met2}
\subsubsection{Original Network}
The original reaction network, containing 645 species and 6163 reactions, came from \citet{Fogel2011}, which was based on the Ohio State University gas-phase model from \citet{Smith2004}. This network has since been updated \citep[][and references therein]{Anderson2021}.

Reaction types considered in the model include gas-phase reactions, grain-surface reactions, photoreactions, cosmic-ray ionization, X-ray-induced UV photolysis, and X-ray ionization of H$_2$ and He. Every molecular species has an adsorption,  a desorption, a photodesorption, and a cosmic-ray desorption process in the network. The network also includes self-shielding of CO, H$_2$, and N$_2$.

\subsubsection{Updated Network}
We modify our existing reaction network guided by the reactions presented in \citet{Laas2019paper}. \citet{Laas2019paper} modeled sulfur chemistry in interstellar clouds using an updated network that includes many new reactions, such as those described in \citet{Vidal2017}. These reactions include several gas-phase reactions, grain-surface reactions, photoreactions, and cosmic-ray ionization. 
We specifically update the reactions in our network that appear in \citet{Laas2019paper} and add new reactions that do not appear in our network. Existing reactions that do not include sulfur are left unchanged from \citet{Anderson2021}. The updated network includes 707 species and 6825 reactions.

We confirm that no reactions are repeated, all reactions conserve charge and mass, and every molecular species has a destruction and production reaction. We also compared the output of our new network with that of the original network to confirm that the abundances of key, highly abundant species like CO and H$_2$O were minimally affected by the changes.

\subsection{Binding Energy} \label{met3}
Binding energies are an important parameter in chemical modeling, determining where species will exist in a gas or ice phase. For sulfur-bearing species, especially, there is a range of binding energies available in the literature \citep[see][]{Perrero2022}. In updating our network, we add several new sulfur species. When adding binding energies for these species, we default to using the binding energies from \citet{Laas2019paper}. The exception is for the sulfur allotropes (S$_2$ through S$_8$), where we use the binding energies from \citet{Cazaux2022}, though we note the range of sulfur allotrope binding energies in the literature \citep[e.g.,][]{Laas2019paper,Ligterink2023}. We choose to use the values from \citet{Cazaux2022}, as they were obtained in an experiment that specifically focused on sulfur allotrope formation. Lastly, for species that were already present in our network, the only binding energies that we update are for the carbon-sulfur chains (i.e., C$_x$S$_y$) \citep{Laas2019paper} and OCS \citep{Collings2004}. Table \ref{Table2_BE} lists the updates made to binding energies for sulfur-bearing species that were already present in our network. 

\begin{table}[]
\centering
\caption{Updated Binding Energies.}
\label{Table2_BE} 
\begin{tabular}{ccc} 
\hline
\multicolumn{1}{c}{Molecule} & \multicolumn{1}{c}{Original BE (K)} & \multicolumn{1}{c}{Updated BE (K)}\\
\hline
{C$_2$S} & {4180}  & {1080}\\
{C$_3$S} & {5010}  & {1880}\\
{C$_4$S} & {5850}  & {2680}\\
{S$_2$} & {3340}  & {3490}\\
{OCS} & {5270}  & {2888}\\
\hline
\end{tabular}
\end{table}

\subsection{Disk Physical Structure} \label{met4}

\begin{figure*}
\begin{center}
\includegraphics[width=.99\textwidth]{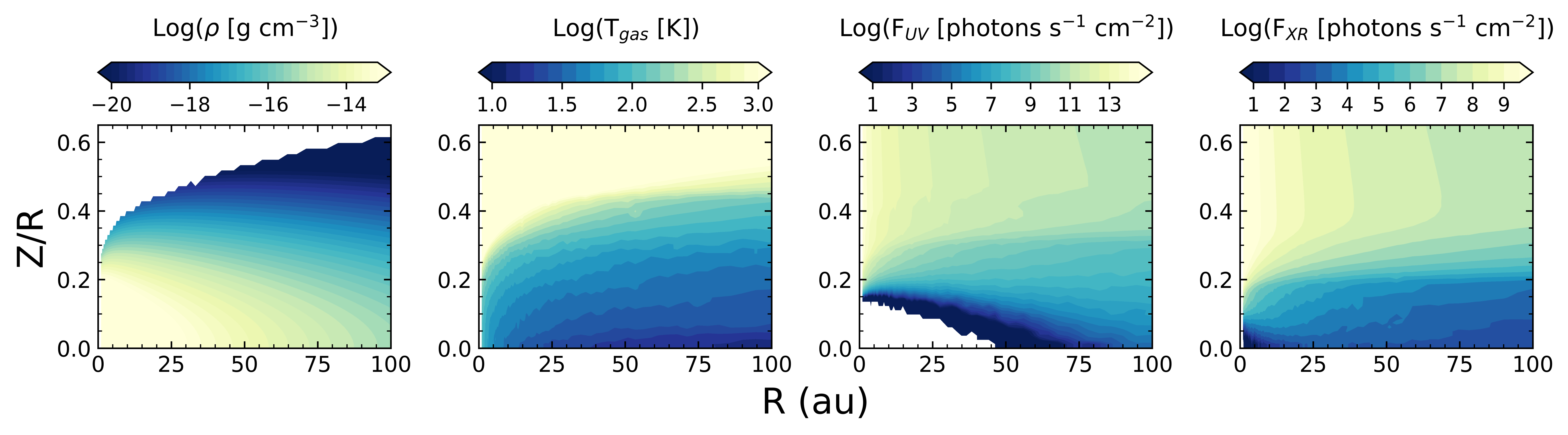}
\caption{Physical environment of the 2D model, showing gas density, gas temperature, ultraviolet radiation, and X-ray radiation.}
\label{fig:phys}
\end{center}
\end{figure*}

Our disk physical conditions come from the 2D disk model presented in \citet{Anderson2021}. We use the fiducial model, where the disk is azimuthally symmetric with a radius of 100 au and a disk mass of 0.01 M$_{\odot}$. At the center of the disk is a 1 M$_{\odot}$ T Tauri star (radius $= 2.8$ R$_{\odot}$, effective temperature $=4300$ K). The physical structure is shown in Figure \ref{fig:phys}.

The dust temperatures assume passive heating by the star and are calculated using TORUS \citep{Harries2004}. The dust consists of two populations, a large and a small dust population, both of which have a minimum dust size of 0.005 $\mu$m. The large population, comprising 90$\%$ of the total dust mass, has a maximum size of 1 mm. The small population has a maximum size of 1 $\mu$m and contains the remaining 10$\%$ of dust mass.

The UV radiation field uses the spectrum from TW Hya, and the X-ray radiation field assumes a total X-ray luminosity of $10^{29.5}$ erg s$^{-1}$ between 1 and 20 keV. These inputs are used in a Monte Carlo radiative transfer code \citep{Bethell2011a,Bethell2011b}. The gas temperatures come from the local UV flux and gas density of the \citet{Bruderer2013} thermochemical models \citep[see][]{Cleeves2015co}. The cosmic-ray ionization rates are presented in \citet{Cleeves2015co}; solar system minimum (SSM) has an incident rate of $1.1\times10^{-18}$ s$^{-1}$ on the surface of the disk, while \citet{Webber1998} (hereafter W98) has an incident rate of $2\times10^{-17}$ s$^{-1}$ and matches values for the dense, molecular interstellar medium.

\subsection{Model Parameters}
We explore the effects of volatile sulfur abundances, C/O ratios, initial sulfur molecular form, and cosmic-ray ionization rate on sulfur molecule abundances. Table \ref{run_names} lists the parameters used in each run, along with our naming convention for each run. We vary the abundance of total volatile sulfur (relative to H) from $10^{-9}$ to $10^{-7}$; this range is chosen because it is consistent with previous observations of sulfur depletion in disks \citep[e.g.,][]{Kama2019, LeGal2021, Keyte2024}. In addition, we explore the effect of the C/O ratio, by reducing the abundance of volatile oxygen. For all three abundance cases, we run models with C/O = 0.36, 0.55, 0.85, and 1.4, and for the mid sulfur case, we also consider C/O = 1.0 and 1.2. This range of values includes C/O = 0.55, the value expected in a disk with no oxygen or carbon depletion \citep{Jenkins2009}. C/O ratios of 1.0 and higher have been proposed as an explanation for the high abundance of CS detections and lack of SO detections in disks \citep{Dutrey2011, Semenov2018, LeGal2021, Keyte2023}. 

Motivated by detections of large sulfur-bearing molecules in comets \citep{Calmonte2016, LeRoy2015}, we explore how the initial molecular form of sulfur affects the later abundances available for comets forming in the disk midplane. For this exploration we adopt the mid-sulfur case ($10^{-8}$) and vary the initial composition of species like CS and SO, H$_2$S, S$_8$ ice, and S.  Additionally, sulfur allotropes could be a stable hiding place for sulfur; by starting the sulfur in S$_8$ ice, we can see if and how other detectable gas-phase molecules are affected by this initial form to predict whether this scenario is testable with observations. Lastly, we run one model (mid10) using the W98 cosmic-ray ionization rate, which is about 18 times higher than the SSM rate. The SSM rate ($1.1\times10^{-18}$ s$^{-1}$) is adopted as our ``standard'' based on observations of molecules that are known cosmic-ray tracers in disks are better fit by models with some degree of cosmic-ray reduction; however, the extent and position vary \citep[e.g.,][]{Cleeves2015co,Seifert2021,Long2024}. Thus, we include a higher value, W98 ($2\times10^{-17}$ s$^{-1}$), typical of the dense molecular interstellar medium (ISM), to test the impact of this assumption.

\begin{table}[ht]
\begin{center}
\caption{2D Model Runs.}
\label{run_names} 
\begin{tabular}{ccccc} 
\hline
\multicolumn{1}{c}{Run} & \multicolumn{1}{c}{S/H} & \multicolumn{1}{c}{Initial Form} & \multicolumn{1}{c}{C/O Ratio} & \multicolumn{1}{c}{CR Rate}\\
\hline
{low1} & {$10^{-9}$}  & {CS $\&$ SO} & {0.36} & {SSM}\\
{low2} & {$10^{-9}$}  & {CS $\&$ SO} & {0.55} & {SSM}\\
{low3} & {$10^{-9}$}  & {CS $\&$ SO} & {0.85} & {SSM}\\
{low4} & {$10^{-9}$}  & {CS $\&$ SO} & {1.40} & {SSM}\\
{mid1} & {$10^{-8}$}  & {CS $\&$ SO} & {0.36} & {SSM}\\
{mid2} & {$10^{-8}$}  & {CS $\&$ SO} & {0.55} & {SSM}\\
{mid3} & {$10^{-8}$}  & {CS $\&$ SO} & {0.85} & {SSM}\\
{mid4} & {$10^{-8}$}  & {CS $\&$ SO} & {1.00} & {SSM}\\
{mid5} & {$10^{-8}$}  & {CS $\&$ SO} & {1.20} & {SSM}\\
{mid6} & {$10^{-8}$}  & {CS $\&$ SO} & {1.40} & {SSM}\\
{mid7} & {$10^{-8}$}  & {H$_2$S} & {0.36} & {SSM}\\
{mid8} & {$10^{-8}$}  & {S$_8$ ice} & {0.36} & {SSM}\\
{mid9} & {$10^{-8}$}  & {S} & {0.36} & {SSM}\\
{mid10} & {$10^{-8}$}  & {CS $\&$ SO} & {0.36} & {W98}\\
{high1}   & {$10^{-7}$}  & {CS $\&$ SO} & {0.36} & {SSM}\\
{high2}   & {$10^{-7}$}  & {CS $\&$ SO} & {0.55} & {SSM}\\
{high3}   & {$10^{-7}$}  & {CS $\&$ SO} & {0.85} & {SSM}\\
{high4}   & {$10^{-7}$}  & {CS $\&$ SO} & {1.40} & {SSM}\\
\hline
\end{tabular}
\end{center}
\end{table}

\section{Results} \label{sec:result}

The results of the model runs, using varying initial condition parameters, are presented below. For the names of model runs, refer to Table \ref{run_names}.

\subsection{Standard Case} \label{res2}
\begin{figure*}
\begin{center}
\includegraphics[width=.99\textwidth]{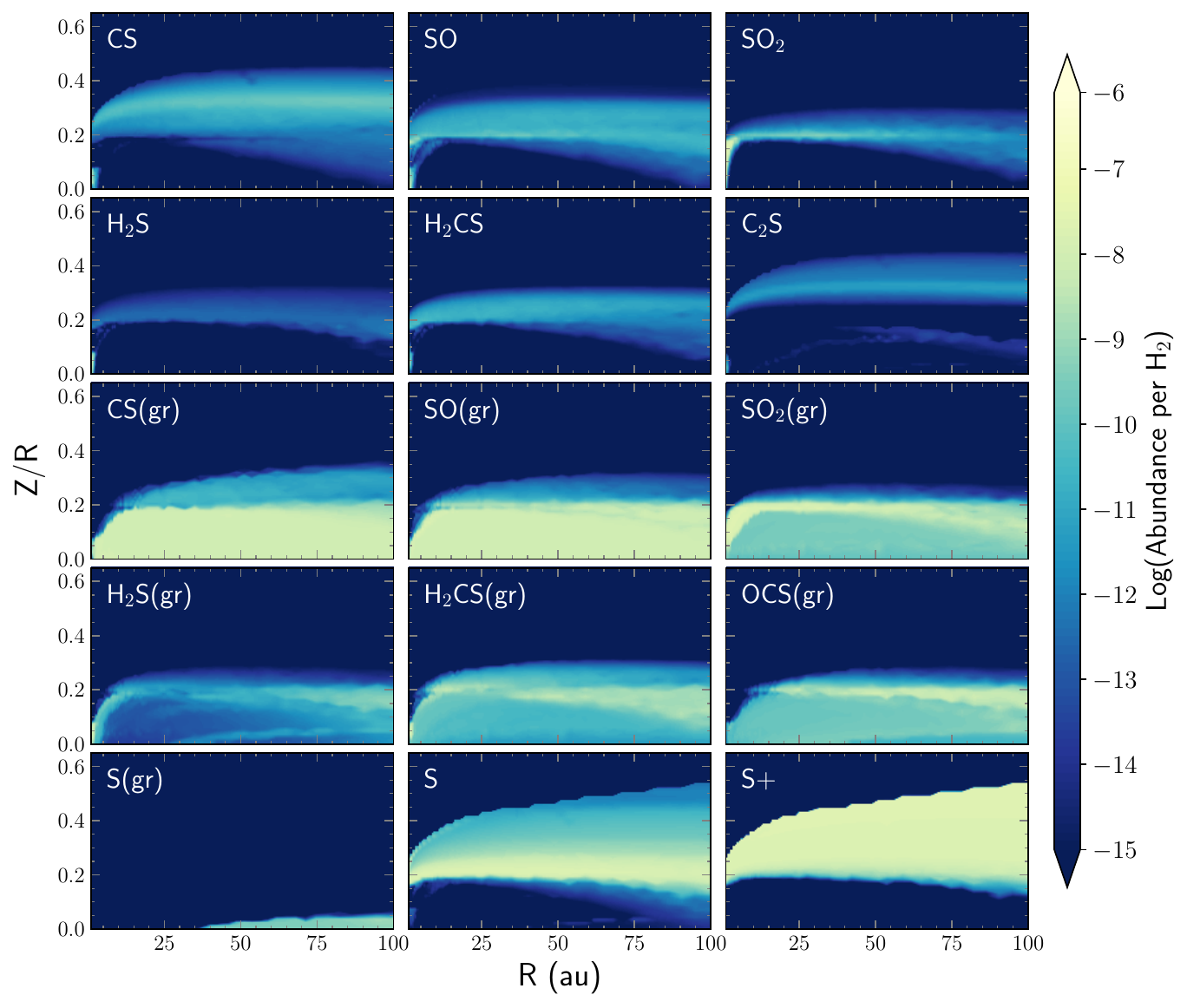}
\caption{Abundances of 15 sulfur-bearing species in a generic disk for the standard case (mid1). The total sulfur abundance is S/H$_2$ $=2\times10^{-8}$, C/O = 0.36, and the initial sulfur form is gas-phase CS and SO.}
\label{fig:low}
\end{center}
\end{figure*}

We refer to the standard case (mid1) as the run where S/H = $10^{-8}$, C/O $=0.36$, and sulfur is initially in gas-phase CS and SO. We use this model as our base case for comparison to other model runs. The abundances of 15 sulfur species are shown in Figure \ref{fig:low}.

\subsection{Effects of Volatile Sulfur Abundance}
\begin{figure}
\begin{center}
\includegraphics[width=0.47\textwidth]{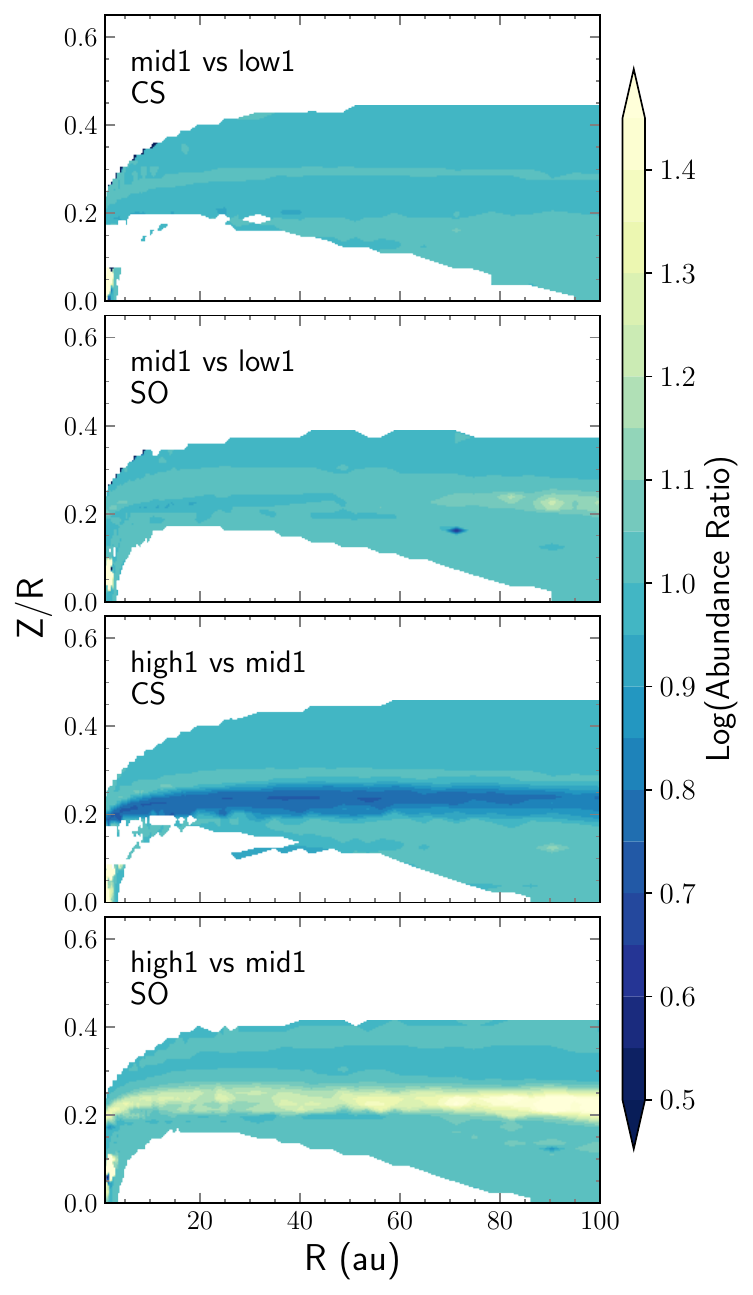}
\caption{Variations in the abundance of CS and SO for different volatile sulfur abundances. The first two plots compare mid1 to low1 (S/H = $10^{-8}$ to $10^{-9}$), and the second two plots compare high1 to mid1 (S/H = $10^{-7}$ to $10^{-8}$). We only plot regions where the abundance of the shown molecule is greater than $10^{-15}$ relative to H$_2$.}
\label{fig:CSSOabund}
\end{center}
\end{figure}

We vary the volatile sulfur abundance relative to H from $10^{-9}$ to $10^{-7}$. Figure \ref{fig:CSSOabund} shows the enhancements in CS and SO as the volatile abundance is increased. We compare the mid case to the low case, and the high case to the mid case. In the mid case, CS and SO are roughly 1 order of magnitude higher in abundance throughout the disk, relative to the low case. In the high case, the enhancement is heterogeneous, with the region around Z/R = 0.2 varying from a uniform increase. CS is enhanced in abundance by less than 1 order of magnitude, while SO increases in abundance by more than 1 order of magnitude.

\subsection{Effects of C/O Ratio}
\begin{figure}
\begin{center}
\includegraphics[width=.47\textwidth]{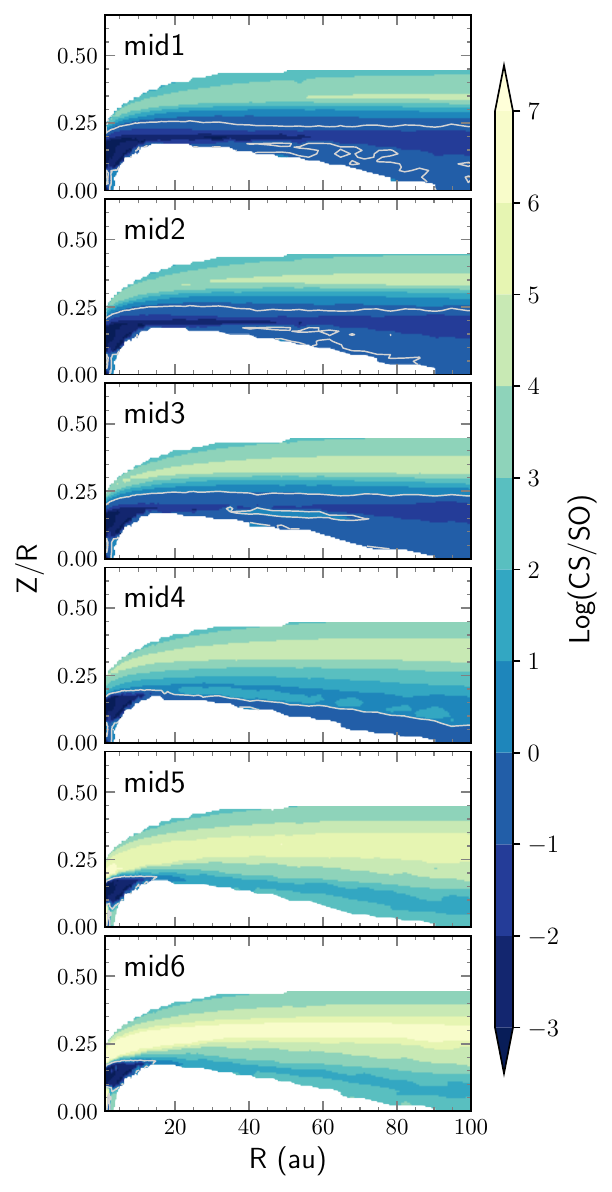}
\caption{Log of the CS/SO ratio as C/O ratio increases from 0.36 to 1.40 (see Table \ref{run_names}). The light gray contour lines show where the CS/SO ratio equals the C/O ratio for that run. For each plot, we only plot regions where the abundance of at least one molecule is greater than $10^{-15}$ relative to H$_2$.}
\label{fig:CSSOCO}
\end{center}
\end{figure}

For each abundance case, we explore the effects of a C/O ratio ranging from 0.36 to 1.40. The C/O ratio was altered by depleting oxygen from H$_2$O ice and CO, leaving excess C in neutral atomic carbon when C/O $>$ 1. Here, we consider the mid sulfur case, though we note that the low and high sulfur cases had similar results. Figure \ref{fig:CSSOCO} shows the CS/SO ratio as the C/O ratio increases. While the C/O ratio increases from only 0.36 to 1.40, the CS/SO ratio spans several orders of magnitude. The solid contour lines show where the CS/SO ratio equals the C/O ratio for each run. As the C/O ratio increases, this contour line moves lower in the disk.

\subsection{Effects of Initial Sulfur Form}
\begin{figure*}
\begin{center}
\includegraphics[width=.99\textwidth]{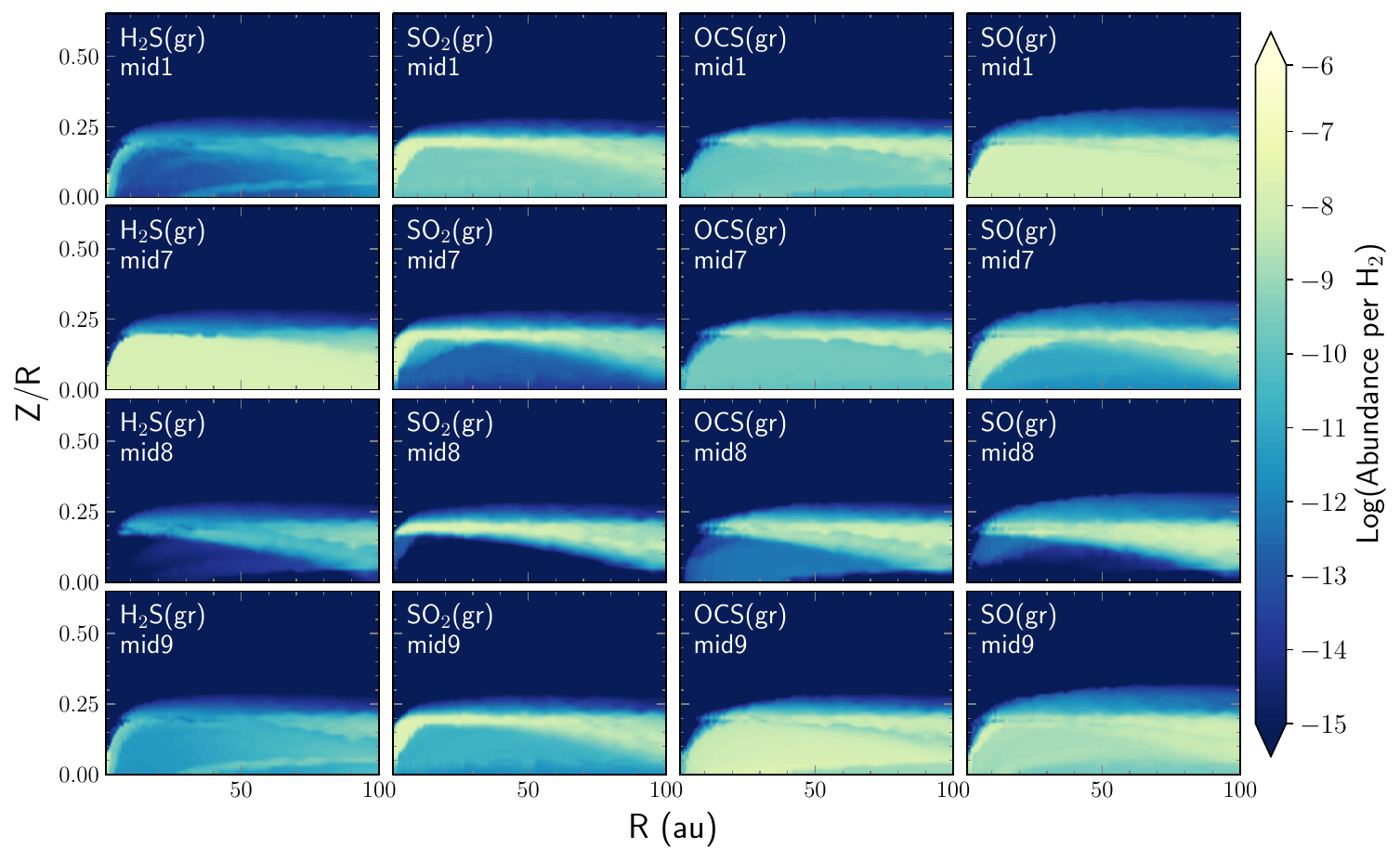}
\caption{Abundances of four ice species, H$_2$S, SO$_2$, OCS, and SO, for the four models where we varied the initial sulfur form (mid1: CS and SO, mid7: H$_2$S, mid8: S$_8$ ice, and mid9: S). For more specifics on the model runs, refer to Table \ref{run_names}.}
\label{fig:comet}
\end{center}
\end{figure*}

For most of our models, we start the sulfur in gas-phase CS and SO. For the mid sulfur case, we also run models with the sulfur starting in H$_2$S, S$_8$ ice, and S. In these three cases (H$_2$S, S$_8$ ice, and S), the gas-phase abundances (above Z/R $\approx0.2$) remain roughly the same as when sulfur starts in CS and SO, while the ice-phase abundances reflect the initial sulfur form. The one exception to this latter finding is the case where the sulfur begins as atomic S; in this case, the main sulfur form in the ice phase is OCS ice. The abundances of several ice species, compared between these four models, are shown in Figure \ref{fig:comet}.

\subsection{Effects of Cosmic-Ray Ionization Rate}

Increasing the cosmic-ray ionization rate has very little effect on the resulting abundances. The W98 rate causes a slight enhancement in H$_2$S ice and H$_2$CS ice in the outer regions of the disk, close to the midplane (not shown).

\section{Discussion} \label{res4}

Here, we discuss in more depth the results and their implications. We consider separately the effects in the warm upper layers of the disk (Z/R $= 0.2-0.4$), where gas-phase molecules dominate, and in the cold lower layers of the disk (Z/R $\le0.2$), where ices dominate. Above Z/R $\sim0.4$, gas-phase molecules do not survive, and S$^+$ dominates.

\subsection{Upper Layers (Z/R = 0.2-0.4)} \label{disc1}

In the warm upper layers of the disk (Z/R $\approx 0.2-0.4$), we find the majority of sulfur in gas-phase molecules. The abundance of CS peaks at Z/R $\approx 0.3$, while SO peaks at Z/R $\approx 0.2$. The initial sulfur form does not have a noticeable effect on the abundances of the gas-phase sulfur species in the disk upper layers. This behavior is a result of the relatively warm temperatures and high radiation field, where any initially complex molecules are soon destroyed.

\subsubsection{Column Densities}\label{columndensity}
\begin{figure*}
\begin{center}
\includegraphics[width=.99\textwidth]{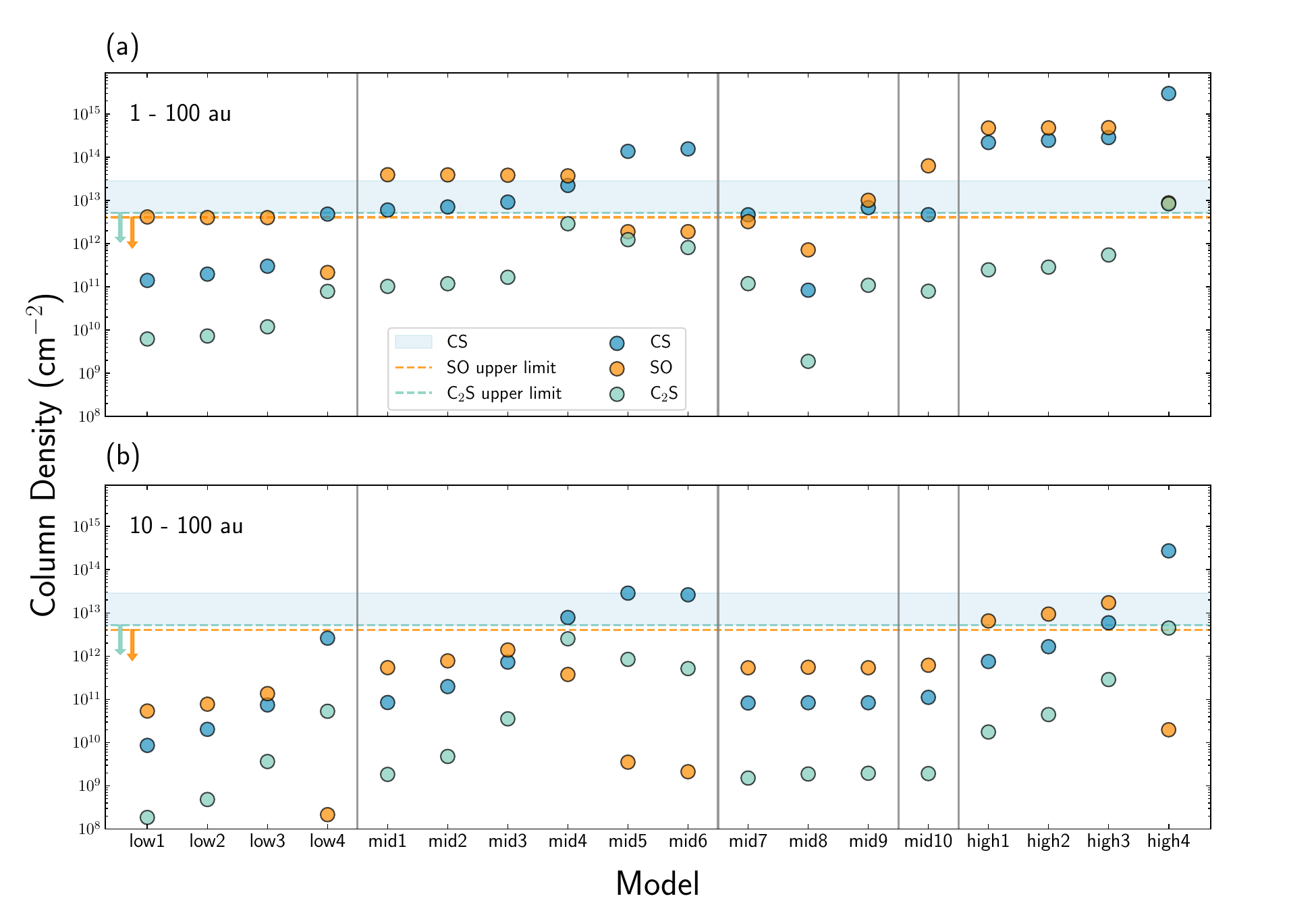}
\caption{Disk-averaged column densities of CS, SO, and C$_2$S in our models, compared to observational data. The top plot is the average from 1 to 100 au, and the bottom plot is the average from 10 to 100 au. Model runs are listed in Table \ref{run_names}. The blue shaded region shows the 1-sigma detected CS column densities from \citet{LeGal2021}, and the dashed orange and mint green lines show the upper limits of SO and C$_2$S, respectively, also from \citet{LeGal2021}. These upper limits are the values for the disk around IM Lup, chosen because IM Lup is the most similar in age and stellar mass to our modeled disk.}
\label{fig:cd_avg}
\end{center}
\end{figure*}

\citet{LeGal2021} reported the observed column density of CS in 10 disks (8 T Tauri and 2 Herbig Ae stars), which varies from about $5.3 \times 10^{12}$ to $2.9 \times 10^{13}$ cm$^{-2}$, as well as upper limits on SO and C$_2$S for 5 of these disks. Figure \ref{fig:cd_avg} shows our disk-averaged column densities of CS, SO, and C$_2$S for all model runs (refer to Table \ref{run_names} for run parameters). In Figure \ref{fig:cd_avg}a, the average is calculated from 1 to 100 au. In Figure \ref{fig:cd_avg}b, the average is calculated from a radius of 10 to 100 au. We also show the values from \citet{LeGal2021} for comparison.

The motivation for excluding the inner 10 au from Figure \ref{fig:cd_avg}b is that our models show a high abundance in column density of both CS and SO within 10 au of the star (see Appendix \ref{appendix}), which skews the average. Observations of CS in disks do not show a spike in column density close to the star, and in many cases, there is a decrease in CS emission at small radii, resulting in an annulus of emission \citep{LeGal2021, Teague2018a}. This peak not being seen in observations could be due to physical effects or model limitations. For example, the inner disk dust opacity, even at millimeter wavelengths, is known to be quite high \citep[e.g.,][]{Huang2018} and so molecular emission can be hidden behind optically thick dust. It is also possible that this peak arises from simplifications in the model, such as having a static physical structure without, e.g., grain growth, or an insufficiently complex chemical network. These model simplifications likely only affect the portion of the disk closest to the star, where the densities are the highest. In addition, comparing column densities does not factor in optical depth effects, either from the line itself or how radial temperature gradients may impact excitation of a given rotational transition. Thus, we opt to compare the outer disk to observations in this work. 

As can be seen in Figure \ref{fig:cd_avg}b, increasing the C/O ratio for a fixed physical structure has a large effect on gas-phase abundances (broadly, resulting in increased CS and C$_2$S and decreased SO), consistent with past models and observations \citep{LeGal2021, Keyte2023, Keyte2024}. While the C/O ratio only varies by a factor of $\approx4$ (from 0.36 to 1.40), the column density of CS increases by more than 2 orders of magnitude, and that of SO decreases by over 2 orders of magnitude. This trend remains the same for all volatile sulfur abundance cases that we tried (low, mid, and high). Interestingly, SO increases slightly in column density from C/O = 0.36 to 0.85; this slight increase corresponds to a decrease in SO$_2$ column density (not shown). We also see that the initial sulfur form (mid7, mid8, and mid9) has a negligible effect on the column densities of CS, SO, and C$_2$S, and an increased cosmic-ray ionization rate (mid10) results in a slight increase in CS column density.

Given the large variations in molecular sulfur column densities for our different C/O compositions, it is natural to ask if these changes are driven by C/O or instead by other parameters of our models - chemical or physical. Specifically, \citet{Wakelam2004} found that in their models of sulfur chemistry in hot cores, the ratios of sulfur-bearing species are strongly affected by temperature and density. If we cross-compare the range of temperatures and densities considered by \citet{Wakelam2004}, 100 - 300 K and $10^{5}$ -~$10^{7}$ cm$^{-3}$ respectively, to our disk model's conditions, there is greatest overlap at Z/R$~\approx 0.2-0.4$ and R$~\approx 20-80$ au. If we examine the physical conditions traced out along a line fixed at Z/R$~= 0.3$ (e.g., Figure \ref{fig:phys}) with the CS/SO ratio along this same line (Figure \ref{fig:CSSOabund}), the CS/SO ratio is roughly constant despite density decreases by over 1 order of magnitude, temperature decreases by about 250 K, and both UV and X-Ray fields decreasing by over 3 orders of magnitude. Along this same line, for models with C/O ratio increasing from 0.36 to 1.4 (mid1 to mid6), the corresponding CS/SO ratio increases by $\sim 2-3$ orders of magnitude. While this is only one disk structure, we find that the underlying C/O ratio has the strongest effect on the CS/SO column density ratio. The difference between these models and the hot core models is likely due to the different radiation field (UV and X-ray) conditions. Nonetheless, a larger parameter space exploration of different physical assumptions would be beneficial to explore the magnitude of the impact on CS/SO introduced by different disk physical structures. 

How do our column densities compare to observed values? The prevalence of CS observed in disks and the lack of SO detections suggest that CS is higher in abundance than SO \citep{LeGal2021}. Out of all of our models, the mid sulfur cases (S/H = $10^{-8}$) with C/O $\geq 1.0$  best match the column densities reported in \citep{LeGal2021}. In these models, CS lies in the range of reported detections, and C$_2$S and SO lie below the upper limits. If we include the inner 10 au in the average (Figure \ref{fig:cd_avg}a), then our results support an intermediary case where S/H = $10^{-9} - 10^{-8}$ and the C/O ratio is $> 1.0$. Whether or not we include the inner 10 au, our results are consistent with a large fraction of sulfur being in refractory form \citep{Kama2019}. \citet{Keyte2024} found through modeling of observations that S/H $\sim10^{-8}$, in agreement with our findings.

The sample studied in \citet{LeGal2021} includes 10 disks, including 5 disks that were studied in \citet{LeGal2019}. One of those disks, DM Tau, was also studied in \citet{Semenov2018} and in \citet{Dutrey2011}, who calculated a CS column density of $2-6\times10^{12}$ cm$^{-2}$ and $3.5 \pm 0.1 \times10^{12}$ cm$^{-2}$, respectively. The CS column densities have been calculated for other disks, as well. \citet{Dutrey2011} determined the CS column density in GO Tau to be $2.0 \pm 0.16 \times10^{12}$ cm$^{-2}$, and \citet{Phuong2021} calculated a CS column density in GG Tau of $2.2 \times10^{13}$ cm$^{-2}$. All of these values are in agreement with or just below the range in \citet{LeGal2021}. Comparing these additional observations to Figure \ref{fig:cd_avg}, the models that best agree are C/O ratio greater than 0.85, and S/H = $10^{-9} - 10^{-8}$.

A high C/O ratio suggests that volatile oxygen is depleted in disks relative to interstellar volatile C and O abundances, leading to a lack of oxygen available for gas-phase molecules. Previous observations of disks have shown that volatile oxygen is in low abundance \citep[e.g.,][]{Dutrey1994, Ansdell2016, Long2017, Du2017}, raising questions as to how oxygen can be depleted from the volatile reservoir. One mechanism through which this depletion can occur is H$_2$O freezing on dust grains and then settling to the midplane, where the oxygen remains trapped as ice \citep{Hogerheijde2011, Bergin2016, Du2017}. This mechanism would result in a high C/O ratio in the upper layers of the disk and a low C/O ratio in the lower layers of the disk, where the H$_2$O settles. We note that our models do not take into account this settling mechanism, as there is no vertical transport of matter in our models.  However, given that our models show that molecular sulfur probes intermediate vertical heights, we can interpret our model results as pointing to an elevated C/O ratio in the upper disk layers.

\subsubsection{Gas-phase Observations}

The most abundant gas-phase species in our models vary slightly with the C/O ratio. For example, at Z/R = 0.3 and R = 40 au, as the C/O ratio increases, SO and HS are replaced by C$_2$S and C$_3$S as two of the most abundant sulfur carriers. Based on our model output, after CS, C$_2$S should be the most abundant sulfur molecule in disks. Looking at Figure \ref{fig:cd_avg}, in the mid4 model, which is one of the closest matches to observational data, C$_2$S is just shy of the upper limits on observations from \citet{LeGal2021}. Future deep observations targeting C$_2$S would provide observational support that a high C/O ratio is present in the upper layers of disks.

\subsection{Lower Layer (Z/R $\leq$ 0.2)}

Lower in the disk, where temperatures are colder, initial sulfur form plays a big role. In the three molecular cases we try (initial sulfur as gas-phase CS and SO, as gas-phase H$_2$S, or as S$_8$ ice), the main sulfur carriers at low Z/R are the ice-phase versions of the initial carriers. When sulfur starts as atomic S, the main carrier is OCS ice, followed by S ice. These results suggest that the initial sulfur form in a protoplanetary disk will largely determine the ices available for comet and planet formation, as sulfur molecules are minimally processed in the midplane in our models.

\subsubsection{Allotropes}

\begin{figure}
\begin{center}
\includegraphics[width=.47\textwidth]{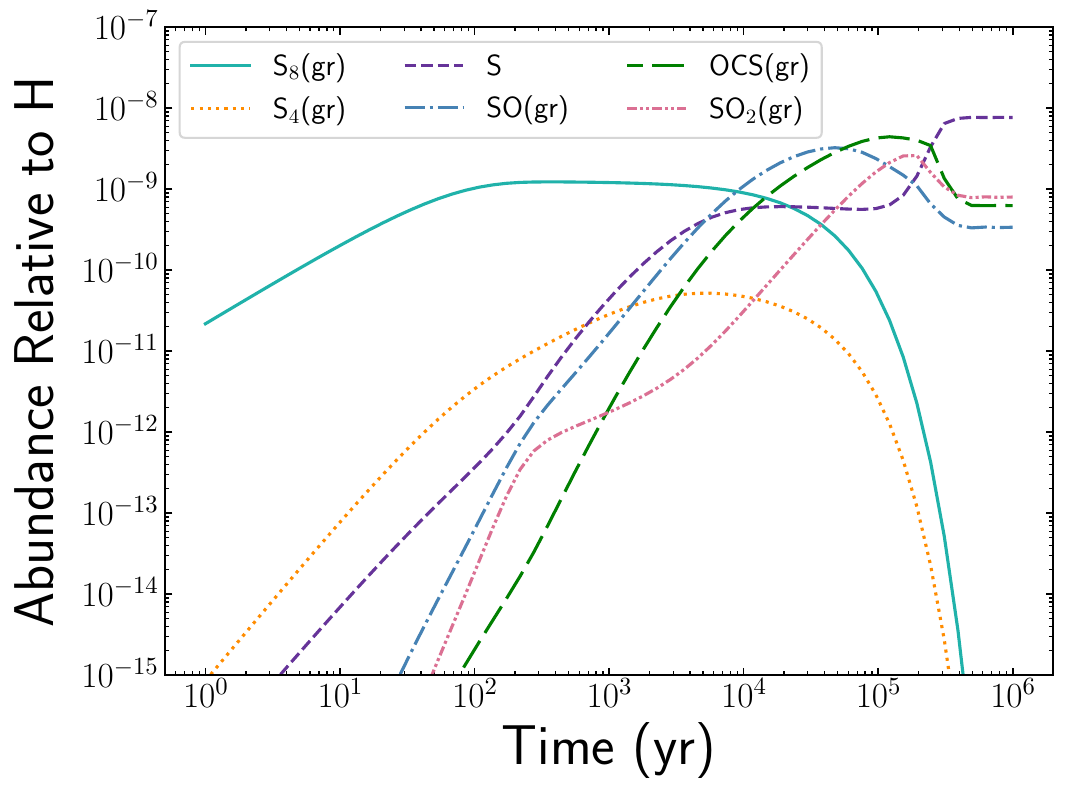}
\caption{Abundance of sulfur species over time at R $= 40$ au, Z $= 8$ au (Z/R $= 0.2$), for run mid8, where sulfur begins in the form S$_8$ ice. Sulfur allotropes are destroyed around $10^{4}-10^{5}$ yr, as simpler sulfur forms (S, SO$_2$ ice, OCS ice, and SO ice) become the dominant sulfur carriers.}
\label{fig:S8time}
\end{center}
\end{figure}

Sulfur is capable of forming large sulfur chains, S$_2$ through S$_8$, and one possible hiding spot for sulfur is in these large, stable allotropes. In our models,  which have limited allotrope chemistry, there is no appreciable build-up of S$_3$ or higher, in disagreement with this hypothesis. The only model in which we see significant amounts of allotropes is when sulfur begins as S$_8$ ice, where it remains in that form in the lower layers of the disk, but is broken apart in the upper layers. Figure \ref{fig:S8time} shows the abundance of several sulfur molecules over time at a disk point where the S$_8$ ice is broken down (R = 40 au, Z = 8 au; Z/R = 0.2), which has a gas and dust temperature of 47~K. S$_8$ ice declines sharply at about $5 \times 10^4$ yr, broken down by photons into smaller sulfur allotrope ices, which are further broken down into S ice. S ice thermally desorbs from the grain and is then available for gas-phase reactions.

Our models suggest that S$_8$ is not a main carrier of sulfur, unless sulfur begins in that form. \citet{Laas2019paper} did not find significant build-up of allotropes in their molecular cloud models, either. More recently, \citet{Shingledecker2020} incorporated cosmic-ray-driven radiation chemistry and nondiffusive bulk reactions into their dense molecular cloud models. They also based their network on the one presented in \citet{Laas2019paper}, and by incorporating these two additional processes, they found a significant buildup of sulfur allotropes, along with OCS and SO$_2$ ices. If sulfur allotropes do indeed form in dense molecular clouds and are inherited by the disk, our models suggest that they will be retained in the midplane of the disk. Unfortunately, S$_8$ does not have any strong infrared active modes and therefore cannot be detected in disks or clouds \citep{Palumbo1997}. We would need observations of other ices in disks to rule out the possibility that S$_8$ ice is a main carrier.

\subsubsection{Implications for Comet Formation}
Our results have interesting implications for comets, which form in the midplane of the disk and are composed of an icy nucleus. Comets are thought to remain largely unprocessed since their formation, which means that their compositions are representative of the conditions present in the solar nebula \citep{Calmonte2016}. Several sulfur species have been detected in comets, and sulfur is not observed to be depleted \citep{Calmonte2016}. For example, in the coma of comet 67P/Churyumov-Gerasimenko, the ROSINA instrument \citep{Balsiger2007} on the European Space Agency's Rosetta spacecraft detected H$_2$S, SO$_2$, SO, and OCS, as well as several other less abundant sulfur species including S$_3$ and S$_4$ \citep{Calmonte2016, LeRoy2015}. H$_2$S comprises the majority of the sulfur budget, accounting for $57\%$ of the sulfur. SO$_2$, SO, and OCS together make up about $14\%$ of the budget. 

Comparing our model ice abundances to detections in comets, the diversity of ices produced in our generic model (where sulfur starts as CS and SO) produces high abundances of SO, OCS, and SO$_2$ ices, but low abundances of H$_2$S ice. However, in our model in which sulfur initially begins as H$_2$S, we do get a high abundance of H$_2$S ice in the lower layers of the disk, which suggests that the H$_2$S present in comets was present before the disk formed. Recently, \citet{Bariosco2024} presented a computational study of H$_2$S binding energies  on amorphous water ice and obtained a wide range of values, 57  - 2406 K. This range is lower than the binding energy we assume (3660~K) and for the low end of this range would imply no H$_2$S freeze-out at any location in our disk model. Given the wide spread of values for this parameter, follow-up on the impact of the H$_2$S binding energy on sulfur chemistry would be warranted.

Our models produce substantial CS ice, which is inconsistent with the nondetection of CS in the coma of 67P \citet{Calmonte2016}. One possibility is that the midplane has a much lower C/O ratio than present in our models, as H$_2$O freezes and settles to the midplane \citep{Hogerheijde2011, Bergin2016, Du2017}. We did not model a C/O ratio less than 0.36, but it is likely that a higher abundance of oxygen in the midplane would lead to higher abundances of oxygen-bearing sulfur species relative to CS ice.

\subsubsection{Ice-phase Observations}
Sulfur ices have yet to be detected in protoplanetary disks, but OCS ice has been detected in a molecular cloud \citep{McClure2023}. Based on our models, future observations of protoplanetary disks targeting OCS, H$_2$S, SO$_2$, SO, and CS ices would be most likely to yield detections. These observations would be valuable for determining the initial sulfur form in disks and for connecting the protoplanetary disk phase to cometary abundances.

In several ices, including SO$_2$, OCS, and H$_2$CS, we observe that the peak abundance occurs not in the midplane, but at Z/R $\approx0.2$ (see Figures \ref{fig:low} and \ref{fig:comet}). These species have larger binding energies than CS and SO, so they can remain in the ice phase higher in the disk. Observations of disks should take into account that absorption from ices such as SO$_2$, OCS, and H$_2$CS might not originate in the midplane, where comets and planets are forming, but higher in the disk.

\section{Conclusions} \label{sec:conclusions}

We present an updated reaction network for protoplanetary disk chemical modeling that includes several new sulfur species and reactions, including sulfur allotropes. We run a 2D time-dependent disk chemical model \citep{Fogel2011} to study the main sulfur molecules that form, and we explore the effects of varying the volatile sulfur abundance, C/O ratio, initial sulfur molecular form, and cosmic-ray ionization rate. We also compare our model results to derived column densities made from disk observations.

\begin{itemize}
    \item The most abundant sulfur-bearing molecules include CS, SO, C$_2$S, SO$_2$ ice, and OCS ice. 
    \item Gas-phase sulfur abundances, like CS and SO, are greatly affected by the C/O ratio. A C/O ratio $\geq1.0$ and S/H = $10^{-8}$ provide the best match to observational data. This result suggests that the majority of sulfur is in refractory form.
    \item An increased cosmic-ray ionization rate from $1.8\times10^{-18}$ s$^{-1}$ to $2\times10^{-17}$ s$^{-1}$ has a minimal effect on the abundances of sulfur species relative to our standard case.
    \item In the lower layers of the disk, the most abundant ice-phase sulfur species are determined by the initial sulfur molecular form. Gas-phase molecules in the upper layers of the disk are minimally affected by initial sulfur form.
    \item Assuming that comets undergo little to no chemical processing after formation, then to explain the high abundance of H$_2$S in comets \citep{Calmonte2016}, there must be some H$_2$S inherited by the disk.

\end{itemize}

This work is just the start in moving toward a more comprehensive model of sulfur chemistry in protoplanetary disks. For example, there is a large uncertainty in the binding energies for many sulfur-bearing molecules, and even those molecules with literature estimates vary substantially from paper to paper. More experimentally determined binding energies for sulfur species on a variety of binding surfaces would improve our models and conclusions about midplane chemistry.

An interesting future direction would be to study the effects of varied X-ray fields on sulfur chemistry. \citet{Waggoner2022} modeled the effects of X-ray flaring events on chemical abundances in disks and found that flares caused organosulfides, like C$_4$S, to increase in abundance. It would be interesting to see if, in the presence of X-ray flares, our updated network results in a similar buildup of organosulfides. These results would give us a further glimpse as to how sulfur may have been incorporated into planets in our own solar system, ultimately leading to the emergence of life on Earth.

\acknowledgments
The original motivation for this project stemmed from an RCSA Scialog Conference and was supported by Heising-Simons Foundation Grant \#2022-3995. B.J.W. acknowledges support from the Virginia Initiative on Cosmic Origins (VICO). L.I.C. acknowledges support from the David and Lucile Packard Foundation, Research Corporation for Science Advancement Cottrell Fellowship, NASA ATP 80NSSC20K0529, NSF grant no. AST-2205698, and SOFIA Award 09-0183.

%
%
%
\vspace{5mm}
\software{GNU Parallel \citep{GNUParallel}}
%
%

\appendix

\section{Radial Profile of CS, SO, and C$_2$S Column Densities}
\label{appendix}

\begin{figure*}
\begin{center}
\includegraphics[width=.99\textwidth]{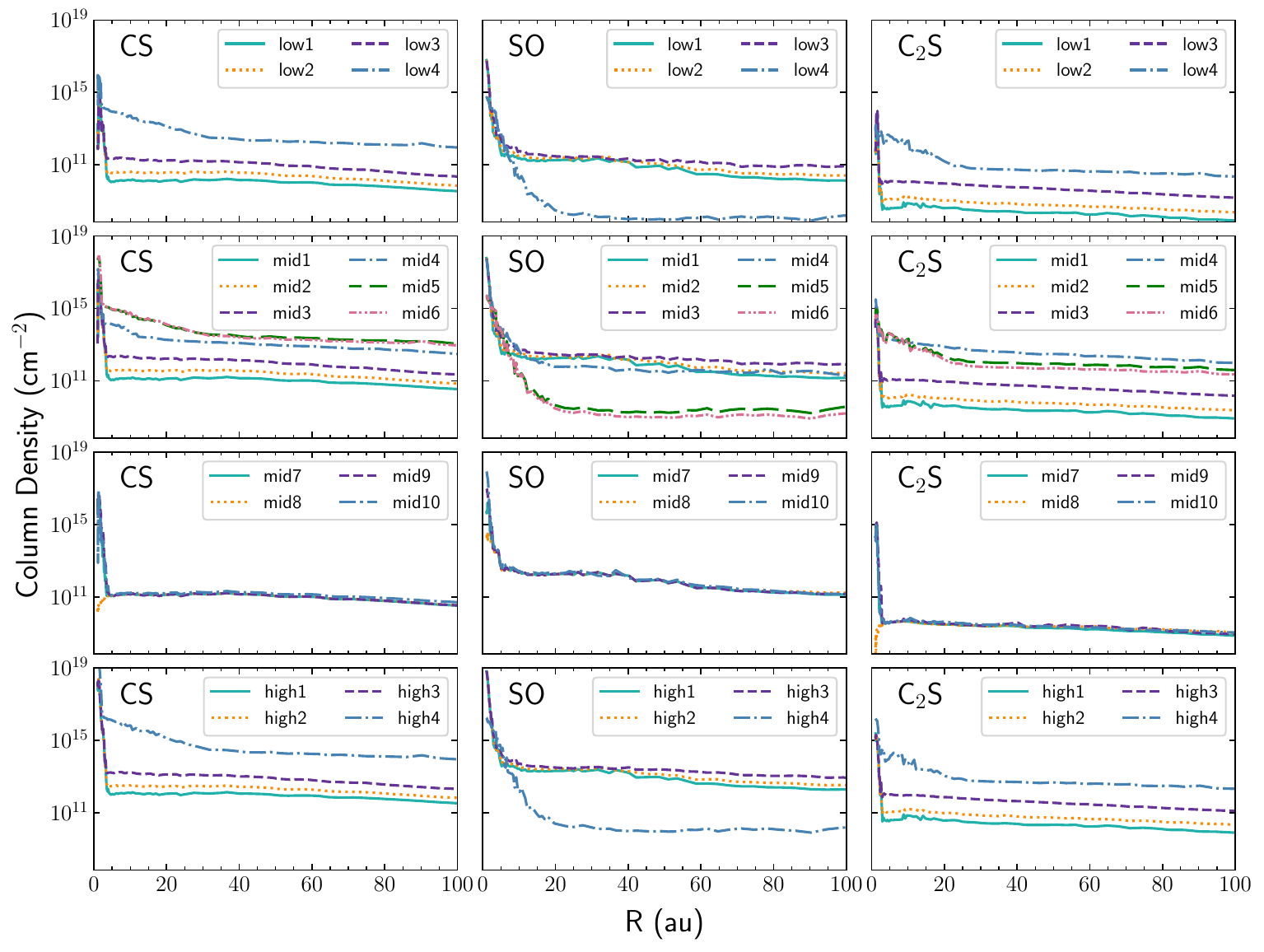}
\caption{Column density as a function of radius for all model runs. Each column is a different molecule, and each row is a different set of model runs. See Table \ref{run_names} for the parameters of each run.}
\label{fig:cd}
\end{center}
\end{figure*}

Figure \ref{fig:cd} shows the radial profiles of the column densities of CS, SO, and C$_2$S for all model runs, motivated by the observations presented in \citet{LeGal2021}. In most runs, all three molecules spike in column density close to the star, at radii less than 10 au. The one exception is that CS and C$_2$S decrease slightly in column density close to the central star in run mid8, where sulfur started as S$_8$ ice. The disk-averaged values are shown in Figure \ref{fig:cd} and discussed in Section \ref{columndensity}.

\begin{thebibliography}{}
\expandafter\ifx\csname natexlab\endcsname\relax\def\natexlab#1{#1}\fi
\providecommand{\url}[1]{\href{#1}{#1}}
\providecommand{\dodoi}[1]{doi:~\href{http://doi.org/#1}{\nolinkurl{#1}}}
\providecommand{\doeprint}[1]{\href{http://ascl.net/#1}{\nolinkurl{http://ascl.net/#1}}}
\providecommand{\doarXiv}[1]{\href{https://arxiv.org/abs/#1}{\nolinkurl{https://arxiv.org/abs/#1}}}

\bibitem[{{Anderson} {et~al.}(2021){Anderson}, {Blake}, {Cleeves}, {Bergin}, {Zhang}, {Schwarz}, {Salyk}, \& {Bosman}}]{Anderson2021}
{Anderson}, D.~E., {Blake}, G.~A., {Cleeves}, L.~I., {et~al.} 2021, \apj, 909, 55, \dodoi{10.3847/1538-4357/abd9c1}

\bibitem[{{Ansdell} {et~al.}(2016){Ansdell}, {Williams}, {van der Marel}, {Carpenter}, {Guidi}, {Hogerheijde}, {Mathews}, {Manara}, {Miotello}, {Natta}, {Oliveira}, {Tazzari}, {Testi}, {van Dishoeck}, \& {van Terwisga}}]{Ansdell2016}
{Ansdell}, M., {Williams}, J.~P., {van der Marel}, N., {et~al.} 2016, \apj, 828, 46, \dodoi{10.3847/0004-637X/828/1/46}

\bibitem[{{Asplund} {et~al.}(2021){Asplund}, {Amarsi}, \& {Grevesse}}]{Asplund2021}
{Asplund}, M., {Amarsi}, A.~M., \& {Grevesse}, N. 2021, \aap, 653, A141, \dodoi{10.1051/0004-6361/202140445}

\bibitem[{{Balsiger} {et~al.}(2007){Balsiger}, {Altwegg}, {Bochsler}, {Eberhardt}, {Fischer}, {Graf}, {J{\"a}ckel}, {Kopp}, {Langer}, {Mildner}, {M{\"u}ller}, {Riesen}, {Rubin}, {Scherer}, {Wurz}, {W{\"u}thrich}, {Arijs}, {Delanoye}, {de Keyser}, {Neefs}, {Nevejans}, {R{\`e}me}, {Aoustin}, {Mazelle}, {M{\'e}dale}, {Sauvaud}, {Berthelier}, {Bertaux}, {Duvet}, {Illiano}, {Fuselier}, {Ghielmetti}, {Magoncelli}, {Shelley}, {Korth}, {Heerlein}, {Lauche}, {Livi}, {Loose}, {Mall}, {Wilken}, {Gliem}, {Fiethe}, {Gombosi}, {Block}, {Carignan}, {Fisk}, {Waite}, {Young}, \& {Wollnik}}]{Balsiger2007}
{Balsiger}, H., {Altwegg}, K., {Bochsler}, P., {et~al.} 2007, \ssr, 128, 745, \dodoi{10.1007/s11214-006-8335-3}

\bibitem[{{Bariosco} {et~al.}(2024){Bariosco}, {Pantaleone}, {Ceccarelli}, {Rimola}, {Balucani}, {Corno}, \& {Ugliengo}}]{Bariosco2024}
{Bariosco}, V., {Pantaleone}, S., {Ceccarelli}, C., {et~al.} 2024, \mnras, 531, 1371, \dodoi{10.1093/mnras/stae1210}

\bibitem[{{Bergin} {et~al.}(2016){Bergin}, {Du}, {Cleeves}, {Blake}, {Schwarz}, {Visser}, \& {Zhang}}]{Bergin2016}
{Bergin}, E.~A., {Du}, F., {Cleeves}, L.~I., {et~al.} 2016, \apj, 831, 101, \dodoi{10.3847/0004-637X/831/1/101}

\bibitem[{{Bethell} \& {Bergin}(2011{\natexlab{a}})}]{Bethell2011a}
{Bethell}, T.~J., \& {Bergin}, E.~A. 2011{\natexlab{a}}, \apj, 739, 78, \dodoi{10.1088/0004-637X/739/2/78}

\bibitem[{{Bethell} \& {Bergin}(2011{\natexlab{b}})}]{Bethell2011b}
---. 2011{\natexlab{b}}, \apj, 740, 7, \dodoi{10.1088/0004-637X/740/1/7}

\bibitem[{{Booth} {et~al.}(2021){Booth}, {van der Marel}, {Leemker}, {van Dishoeck}, \& {Ohashi}}]{Booth2021}
{Booth}, A.~S., {van der Marel}, N., {Leemker}, M., {van Dishoeck}, E.~F., \& {Ohashi}, S. 2021, \aap, 651, L6, \dodoi{10.1051/0004-6361/202141057}

\bibitem[{Brosnan \& Brosnan(2006)}]{Brosnan2006}
Brosnan, J.~T., \& Brosnan, M.~E. 2006, The Journal of Nutrition, 136, 1636S, \dodoi{https://doi.org/10.1093/jn/136.6.1636S}

\bibitem[{{Bruderer}(2013)}]{Bruderer2013}
{Bruderer}, S. 2013, \aap, 559, A46, \dodoi{10.1051/0004-6361/201321171}

\bibitem[{{Calmonte} {et~al.}(2016){Calmonte}, {Altwegg}, {Balsiger}, {Berthelier}, {Bieler}, {Cessateur}, {Dhooghe}, {van Dishoeck}, {Fiethe}, {Fuselier}, {Gasc}, {Gombosi}, {H{\"a}ssig}, {Le Roy}, {Rubin}, {S{\'e}mon}, {Tzou}, \& {Wampfler}}]{Calmonte2016}
{Calmonte}, U., {Altwegg}, K., {Balsiger}, H., {et~al.} 2016, \mnras, 462, S253, \dodoi{10.1093/mnras/stw2601}

\bibitem[{{Cazaux} {et~al.}(2022){Cazaux}, {Carrascosa}, {Mu{\~n}oz Caro}, {Caselli}, {Fuente}, {Navarro-Almaida}, \& {Rivi{\'e}re-Marichalar}}]{Cazaux2022}
{Cazaux}, S., {Carrascosa}, H., {Mu{\~n}oz Caro}, G.~M., {et~al.} 2022, \aap, 657, A100, \dodoi{10.1051/0004-6361/202141861}

\bibitem[{{Cleeves} {et~al.}(2015){Cleeves}, {Bergin}, {Qi}, {Adams}, \& {{\"O}berg}}]{Cleeves2015co}
{Cleeves}, L.~I., {Bergin}, E.~A., {Qi}, C., {Adams}, F.~C., \& {{\"O}berg}, K.~I. 2015, \apj, 799, 204, \dodoi{10.1088/0004-637X/799/2/204}

\bibitem[{{Cleeves} {et~al.}(2018){Cleeves}, {{\"O}berg}, {Wilner}, {Huang}, {Loomis}, {Andrews}, \& {Guzman}}]{Cleeves2018}
{Cleeves}, L.~I., {{\"O}berg}, K.~I., {Wilner}, D.~J., {et~al.} 2018, \apj, 865, 155, \dodoi{10.3847/1538-4357/aade96}

\bibitem[{{Collings} {et~al.}(2004){Collings}, {Anderson}, {Chen}, {Dever}, {Viti}, {Williams}, \& {McCoustra}}]{Collings2004}
{Collings}, M.~P., {Anderson}, M.~A., {Chen}, R., {et~al.} 2004, \mnras, 354, 1133, \dodoi{10.1111/j.1365-2966.2004.08272.x}

\bibitem[{{Du} {et~al.}(2017){Du}, {Bergin}, {Hogerheijde}, {van Dishoeck}, {Blake}, {Bruderer}, {Cleeves}, {Dominik}, {Fedele}, {Lis}, {Melnick}, {Neufeld}, {Pearson}, \& {Y{\i}ld{\i}z}}]{Du2017}
{Du}, F., {Bergin}, E.~A., {Hogerheijde}, M., {et~al.} 2017, \apj, 842, 98, \dodoi{10.3847/1538-4357/aa70ee}

\bibitem[{{Dutrey} {et~al.}(1994){Dutrey}, {Guilloteau}, \& {Simon}}]{Dutrey1994}
{Dutrey}, A., {Guilloteau}, S., \& {Simon}, M. 1994, \aap, 286, 149

\bibitem[{{Dutrey} {et~al.}(2011){Dutrey}, {Wakelam}, {Boehler}, {Guilloteau}, {Hersant}, {Semenov}, {Chapillon}, {Henning}, {Pi{\'e}tu}, {Launhardt}, {Gueth}, \& {Schreyer}}]{Dutrey2011}
{Dutrey}, A., {Wakelam}, V., {Boehler}, Y., {et~al.} 2011, \aap, 535, A104, \dodoi{10.1051/0004-6361/201116931}

\bibitem[{{Fogel} {et~al.}(2011){Fogel}, {Bethell}, {Bergin}, {Calvet}, \& {Semenov}}]{Fogel2011}
{Fogel}, J. K.~J., {Bethell}, T.~J., {Bergin}, E.~A., {Calvet}, N., \& {Semenov}, D. 2011, \apj, 726, 29, \dodoi{10.1088/0004-637X/726/1/29}

\bibitem[{{Fuente} {et~al.}(2010){Fuente}, {Cernicharo}, {Ag{\'u}ndez}, {Bern{\'e}}, {Goicoechea}, {Alonso-Albi}, \& {Marcelino}}]{Fuente2010}
{Fuente}, A., {Cernicharo}, J., {Ag{\'u}ndez}, M., {et~al.} 2010, \aap, 524, A19, \dodoi{10.1051/0004-6361/201014905}

\bibitem[{{Fuente} {et~al.}(2016){Fuente}, {Cernicharo}, {Roueff}, {Gerin}, {Pety}, {Marcelino}, {Bachiller}, {Lefloch}, {Roncero}, \& {Aguado}}]{Fuente2016}
{Fuente}, A., {Cernicharo}, J., {Roueff}, E., {et~al.} 2016, \aap, 593, A94, \dodoi{10.1051/0004-6361/201628285}

\bibitem[{{Fuente} {et~al.}(2019){Fuente}, {Navarro}, {Caselli}, {Gerin}, {Kramer}, {Roueff}, {Alonso-Albi}, {Bachiller}, {Cazaux}, {Commercon}, {Friesen}, {Garc{\'\i}a-Burillo}, {Giuliano}, {Goicoechea}, {Gratier}, {Hacar}, {Jim{\'e}nez-Serra}, {Kirk}, {Lattanzi}, {Loison}, {Malinen}, {Marcelino}, {Mart{\'\i}n-Dom{\'e}nech}, {Mu{\~n}oz-Caro}, {Pineda}, {Tafalla}, {Tercero}, {Ward-Thompson}, {Trevi{\~n}o-Morales}, {Rivi{\'e}re-Marichalar}, {Roncero}, {Vidal}, \& {Ballester}}]{Fuente2019}
{Fuente}, A., {Navarro}, D.~G., {Caselli}, P., {et~al.} 2019, \aap, 624, A105, \dodoi{10.1051/0004-6361/201834654}

\bibitem[{{Fuente} {et~al.}(2023){Fuente}, {Rivi{\`e}re-Marichalar}, {Beitia-Antero}, {Caselli}, {Wakelam}, {Esplugues}, {Rodr{\'\i}guez-Baras}, {Navarro-Almaida}, {Gerin}, {Kramer}, {Bachiller}, {Goicoechea}, {Jim{\'e}nez-Serra}, {Loison}, {Ivlev}, {Mart{\'\i}n-Dom{\'e}nech}, {Spezzano}, {Roncero}, {Mu{\~n}oz-Caro}, {Cazaux}, \& {Marcelino}}]{Fuente2023}
{Fuente}, A., {Rivi{\`e}re-Marichalar}, P., {Beitia-Antero}, L., {et~al.} 2023, \aap, 670, A114, \dodoi{10.1051/0004-6361/202244843}

\bibitem[{{Harries} {et~al.}(2004){Harries}, {Monnier}, {Symington}, \& {Kurosawa}}]{Harries2004}
{Harries}, T.~J., {Monnier}, J.~D., {Symington}, N.~H., \& {Kurosawa}, R. 2004, \mnras, 350, 565, \dodoi{10.1111/j.1365-2966.2004.07668.x}

\bibitem[{{Hogerheijde} {et~al.}(2011){Hogerheijde}, {Bergin}, {Brinch}, {Cleeves}, {Fogel}, {Blake}, {Dominik}, {Lis}, {Melnick}, {Neufeld}, {Pani{\'c}}, {Pearson}, {Kristensen}, {Y{\i}ld{\i}z}, \& {van Dishoeck}}]{Hogerheijde2011}
{Hogerheijde}, M.~R., {Bergin}, E.~A., {Brinch}, C., {et~al.} 2011, Science, 334, 338, \dodoi{10.1126/science.1208931}

\bibitem[{{Huang} {et~al.}(2018){Huang}, {Andrews}, {Dullemond}, {Isella}, {P{\'e}rez}, {Guzm{\'a}n}, {{\"O}berg}, {Zhu}, {Zhang}, {Bai}, {Benisty}, {Birnstiel}, {Carpenter}, {Hughes}, {Ricci}, {Weaver}, \& {Wilner}}]{Huang2018}
{Huang}, J., {Andrews}, S.~M., {Dullemond}, C.~P., {et~al.} 2018, \apjl, 869, L42, \dodoi{10.3847/2041-8213/aaf740}

\bibitem[{{Jenkins}(2009)}]{Jenkins2009}
{Jenkins}, E.~B. 2009, \apj, 700, 1299, \dodoi{10.1088/0004-637X/700/2/1299}

\bibitem[{{Kama} {et~al.}(2019){Kama}, {Shorttle}, {Jermyn}, {Folsom}, {Furuya}, {Bergin}, {Walsh}, \& {Keller}}]{Kama2019}
{Kama}, M., {Shorttle}, O., {Jermyn}, A.~S., {et~al.} 2019, \apj, 885, 114, \dodoi{10.3847/1538-4357/ab45f8}

\bibitem[{{Kasting} {et~al.}(1989){Kasting}, {Zahnle}, {Pinto}, \& {Young}}]{Kasting1989}
{Kasting}, J.~F., {Zahnle}, K.~J., {Pinto}, J.~P., \& {Young}, A.~T. 1989, Origins of Life and Evolution of the Biosphere, 19, 252, \dodoi{10.1007/BF02388836}

\bibitem[{{Keyte} {et~al.}(2024){Keyte}, {Kama}, {Chuang}, {Cleeves}, {Drozdovskaya}, {Furuya}, {Rawlings}, \& {Shorttle}}]{Keyte2024}
{Keyte}, L., {Kama}, M., {Chuang}, K.-J., {et~al.} 2024, \mnras, 528, 388, \dodoi{10.1093/mnras/stae019}

\bibitem[{{Keyte} {et~al.}(2023){Keyte}, {Kama}, {Booth}, {Bergin}, {Cleeves}, {van Dishoeck}, {Drozdovskaya}, {Furuya}, {Rawlings}, {Shorttle}, \& {Walsh}}]{Keyte2023}
{Keyte}, L., {Kama}, M., {Booth}, A.~S., {et~al.} 2023, Nature Astronomy, 7, 684, \dodoi{10.1038/s41550-023-01951-9}

\bibitem[{Kushkevych {et~al.}(2021)Kushkevych, Procházka, Gajdács, Rittmann, \& Vítězová}]{Kushkevych2021}
Kushkevych, I., Procházka, J., Gajdács, M., Rittmann, S. K.-M., \& Vítězová, M. 2021, International Journal of Molecular Sciences, 22, 6398, \dodoi{10.3390/ijms22126398}

\bibitem[{{Laas} \& {Caselli}(2019)}]{Laas2019paper}
{Laas}, J.~C., \& {Caselli}, P. 2019, \aap, 624, A108, \dodoi{10.1051/0004-6361/201834446}

\bibitem[{{Lake}(1988)}]{Lake1988}
{Lake}, J.~A. 1988, \nat, 331, 184, \dodoi{10.1038/331184a0}

\bibitem[{{Law} {et~al.}(2023){Law}, {Booth}, \& {{\"O}berg}}]{Law2023}
{Law}, C.~J., {Booth}, A.~S., \& {{\"O}berg}, K.~I. 2023, \apjl, 952, L19, \dodoi{10.3847/2041-8213/acdfd0}

\bibitem[{{Le Gal} {et~al.}(2019){Le Gal}, {{\"O}berg}, {Loomis}, {Pegues}, \& {Bergner}}]{LeGal2019}
{Le Gal}, R., {{\"O}berg}, K.~I., {Loomis}, R.~A., {Pegues}, J., \& {Bergner}, J.~B. 2019, \apj, 876, 72, \dodoi{10.3847/1538-4357/ab1416}

\bibitem[{{Le Gal} {et~al.}(2021){Le Gal}, {{\"O}berg}, {Teague}, {Loomis}, {Law}, {Walsh}, {Bergin}, {M{\'e}nard}, {Wilner}, {Andrews}, {Aikawa}, {Booth}, {Cataldi}, {Bergner}, {Bosman}, {Cleeves}, {Czekala}, {Furuya}, {Guzm{\'a}n}, {Huang}, {Ilee}, {Nomura}, {Qi}, {Schwarz}, {Tsukagoshi}, {Yamato}, \& {Zhang}}]{LeGal2021}
{Le Gal}, R., {{\"O}berg}, K.~I., {Teague}, R., {et~al.} 2021, \apjs, 257, 12, \dodoi{10.3847/1538-4365/ac2583}

\bibitem[{{Le Roy} {et~al.}(2015){Le Roy}, {Altwegg}, {Balsiger}, {Berthelier}, {Bieler}, {Briois}, {Calmonte}, {Combi}, {De Keyser}, {Dhooghe}, {Fiethe}, {Fuselier}, {Gasc}, {Gombosi}, {H{\"a}ssig}, {J{\"a}ckel}, {Rubin}, \& {Tzou}}]{LeRoy2015}
{Le Roy}, L., {Altwegg}, K., {Balsiger}, H., {et~al.} 2015, \aap, 583, A1, \dodoi{10.1051/0004-6361/201526450}

\bibitem[{{Ligterink} \& {Minissale}(2023)}]{Ligterink2023}
{Ligterink}, N.~F.~W., \& {Minissale}, M. 2023, \aap, 676, A80, \dodoi{10.1051/0004-6361/202346436}

\bibitem[{{Long} {et~al.}(2024){Long}, {Cleeves}, {Adams}, {Andrews}, {Bergin}, {Guzm{\'a}n}, {Huang}, {Hughes}, {Qi}, {Schwarz}, {Simon}, \& {Wilner}}]{Long2024}
{Long}, D.~E., {Cleeves}, L.~I., {Adams}, F.~C., {et~al.} 2024, \apj, 972, 88, \dodoi{10.3847/1538-4357/ad5c67}

\bibitem[{{Long} {et~al.}(2017){Long}, {Herczeg}, {Pascucci}, {Drabek-Maunder}, {Mohanty}, {Testi}, {Apai}, {Hendler}, {Henning}, {Manara}, \& {Mulders}}]{Long2017}
{Long}, F., {Herczeg}, G.~J., {Pascucci}, I., {et~al.} 2017, \apj, 844, 99, \dodoi{10.3847/1538-4357/aa78fc}

\bibitem[{{McClure} {et~al.}(2023){McClure}, {Rocha}, {Pontoppidan}, {Crouzet}, {Chu}, {Dartois}, {Lamberts}, {Noble}, {Pendleton}, {Perotti}, {Qasim}, {Rachid}, {Smith}, {Sun}, {Beck}, {Boogert}, {Brown}, {Caselli}, {Charnley}, {Cuppen}, {Dickinson}, {Drozdovskaya}, {Egami}, {Erkal}, {Fraser}, {Garrod}, {Harsono}, {Ioppolo}, {Jim{\'e}nez-Serra}, {Jin}, {J{\o}rgensen}, {Kristensen}, {Lis}, {McCoustra}, {McGuire}, {Melnick}, {{\~A}-berg}, {Palumbo}, {Shimonishi}, {Sturm}, {van Dishoeck}, \& {Linnartz}}]{McClure2023}
{McClure}, M.~K., {Rocha}, W.~R.~M., {Pontoppidan}, K.~M., {et~al.} 2023, Nature Astronomy, 7, 431, \dodoi{10.1038/s41550-022-01875-w}

\bibitem[{{Palumbo} {et~al.}(1997){Palumbo}, {Geballe}, \& {Tielens}}]{Palumbo1997}
{Palumbo}, M.~E., {Geballe}, T.~R., \& {Tielens}, A.~G.~G.~M. 1997, \apj, 479, 839, \dodoi{10.1086/303905}

\bibitem[{{Perrero} {et~al.}(2022){Perrero}, {Enrique-Romero}, {Ferrero}, {Ceccarelli}, {Podio}, {Codella}, {Rimola}, \& {Ugliengo}}]{Perrero2022}
{Perrero}, J., {Enrique-Romero}, J., {Ferrero}, S., {et~al.} 2022, \apj, 938, 158, \dodoi{10.3847/1538-4357/ac9278}

\bibitem[{{Phuong} {et~al.}(2021){Phuong}, {Dutrey}, {Chapillon}, {Guilloteau}, {Bary}, {Beck}, {Coutens}, {Denis-Alpizar}, {Di Folco}, {Diep}, {Majumdar}, {Melisse}, {Lee}, {Pietu}, {Stoecklin}, \& {Tang}}]{Phuong2021}
{Phuong}, N.~T., {Dutrey}, A., {Chapillon}, E., {et~al.} 2021, \aap, 653, L5, \dodoi{10.1051/0004-6361/202141881}

\bibitem[{{Rivi{\`e}re-Marichalar} {et~al.}(2022){Rivi{\`e}re-Marichalar}, {Fuente}, {Esplugues}, {Wakelam}, {le Gal}, {Baruteau}, {Ribas}, {Mac{\'\i}as}, {Neri}, \& {Navarro-Almaida}}]{RiviereMarichalar2022}
{Rivi{\`e}re-Marichalar}, P., {Fuente}, A., {Esplugues}, G., {et~al.} 2022, \aap, 665, A61, \dodoi{10.1051/0004-6361/202142906}

\bibitem[{{Rocha} {et~al.}(2023){Rocha}, {Roncero}, {Bulut}, {Zuchowski}, {Navarro-Almaida}, {Fuente}, {Wakelam}, {Loison}, {Roueff}, {Goicoechea}, {Esplugues}, {Beitia-Antero}, {Caselli}, {Lattanzi}, {Pineda}, {Le Gal}, {Rodr{\'\i}guez-Baras}, \& {Riviere-Marichalar}}]{Rocha2023}
{Rocha}, C. M.~R., {Roncero}, O., {Bulut}, N., {et~al.} 2023, \aap, 677, A41, \dodoi{10.1051/0004-6361/202346967}

\bibitem[{{Rustamkulov} {et~al.}(2023){Rustamkulov}, {Sing}, {Mukherjee}, {May}, {Kirk}, {Schlawin}, {Line}, {Piaulet}, {Carter}, {Batalha}, {Goyal}, {L{\'o}pez-Morales}, {Lothringer}, {MacDonald}, {Moran}, {Stevenson}, {Wakeford}, {Espinoza}, {Bean}, {Batalha}, {Benneke}, {Berta-Thompson}, {Crossfield}, {Gao}, {Kreidberg}, {Powell}, {Cubillos}, {Gibson}, {Leconte}, {Molaverdikhani}, {Nikolov}, {Parmentier}, {Roy}, {Taylor}, {Turner}, {Wheatley}, {Aggarwal}, {Ahrer}, {Alam}, {Alderson}, {Allen}, {Banerjee}, {Barat}, {Barrado}, {Barstow}, {Bell}, {Blecic}, {Brande}, {Casewell}, {Changeat}, {Chubb}, {Crouzet}, {Daylan}, {Decin}, {D{\'e}sert}, {Mikal-Evans}, {Feinstein}, {Flagg}, {Fortney}, {Harrington}, {Heng}, {Hong}, {Hu}, {Iro}, {Kataria}, {Kempton}, {Krick}, {Lendl}, {Lillo-Box}, {Louca}, {Lustig-Yaeger}, {Mancini}, {Mansfield}, {Mayne}, {Miguel}, {Morello}, {Ohno}, {Palle}, {Petit dit de la Roche}, {Rackham}, {Radica}, {Ramos-Rosado}, {Redfield}, {Rogers}, {Shkolnik}, {Southworth}, {Teske}, {Tremblin},
  {Tucker}, {Venot}, {Waalkes}, {Welbanks}, {Zhang}, \& {Zieba}}]{Rustamkulov2023}
{Rustamkulov}, Z., {Sing}, D.~K., {Mukherjee}, S., {et~al.} 2023, \nat, 614, 659, \dodoi{10.1038/s41586-022-05677-y}

\bibitem[{{Santos} {et~al.}(2024){Santos}, {Enrique-Romero}, {Lamberts}, {Linnartz}, \& {Chuang}}]{Santos2024}
{Santos}, J.~C., {Enrique-Romero}, J., {Lamberts}, T., {Linnartz}, H., \& {Chuang}, K.-J. 2024, arXiv e-prints, arXiv:2407.09730, \dodoi{10.48550/arXiv.2407.09730}

\bibitem[{{Seifert} {et~al.}(2021){Seifert}, {Cleeves}, {Adams}, \& {Li}}]{Seifert2021}
{Seifert}, R.~A., {Cleeves}, L.~I., {Adams}, F.~C., \& {Li}, Z.-Y. 2021, \apj, 912, 136, \dodoi{10.3847/1538-4357/abf09a}

\bibitem[{{Semenov} {et~al.}(2018){Semenov}, {Favre}, {Fedele}, {Guilloteau}, {Teague}, {Henning}, {Dutrey}, {Chapillon}, {Hersant}, \& {Pi{\'e}tu}}]{Semenov2018}
{Semenov}, D., {Favre}, C., {Fedele}, D., {et~al.} 2018, \aap, 617, A28, \dodoi{10.1051/0004-6361/201832980}

\bibitem[{{Shingledecker} {et~al.}(2020){Shingledecker}, {Lamberts}, {Laas}, {Vasyunin}, {Herbst}, {K{\"a}stner}, \& {Caselli}}]{Shingledecker2020}
{Shingledecker}, C.~N., {Lamberts}, T., {Laas}, J.~C., {et~al.} 2020, \apj, 888, 52, \dodoi{10.3847/1538-4357/ab5360}

\bibitem[{{Smith} {et~al.}(2004){Smith}, {Herbst}, \& {Chang}}]{Smith2004}
{Smith}, I. W.~M., {Herbst}, E., \& {Chang}, Q. 2004, \mnras, 350, 323, \dodoi{10.1111/j.1365-2966.2004.07656.x}

\bibitem[{Tange(2024)}]{GNUParallel}
Tange, O. 2024, GNU Parallel 20240122 ('Frederik X'),  Zenodo, \dodoi{10.5281/zenodo.10558745}

\bibitem[{{Teague} {et~al.}(2018){Teague}, {Henning}, {Guilloteau}, {Bergin}, {Semenov}, {Dutrey}, {Flock}, {Gorti}, \& {Birnstiel}}]{Teague2018a}
{Teague}, R., {Henning}, T., {Guilloteau}, S., {et~al.} 2018, \apj, 864, 133, \dodoi{10.3847/1538-4357/aad80e}

\bibitem[{{Vastel} {et~al.}(2018){Vastel}, {Qu{\'e}nard}, {Le Gal}, {Wakelam}, {Andrianasolo}, {Caselli}, {Vidal}, {Ceccarelli}, {Lefloch}, \& {Bachiller}}]{Vastel2018}
{Vastel}, C., {Qu{\'e}nard}, D., {Le Gal}, R., {et~al.} 2018, \mnras, 478, 5514, \dodoi{10.1093/mnras/sty1336}

\bibitem[{{Vidal} {et~al.}(2017){Vidal}, {Loison}, {Jaziri}, {Ruaud}, {Gratier}, \& {Wakelam}}]{Vidal2017}
{Vidal}, T. H.~G., {Loison}, J.-C., {Jaziri}, A.~Y., {et~al.} 2017, \mnras, 469, 435, \dodoi{10.1093/mnras/stx828}

\bibitem[{{Waggoner} \& {Cleeves}(2022)}]{Waggoner2022}
{Waggoner}, A.~R., \& {Cleeves}, L.~I. 2022, \apj, 928, 46, \dodoi{10.3847/1538-4357/ac549f}

\bibitem[{{Wakelam} {et~al.}(2004){Wakelam}, {Caselli}, {Ceccarelli}, {Herbst}, \& {Castets}}]{Wakelam2004}
{Wakelam}, V., {Caselli}, P., {Ceccarelli}, C., {Herbst}, E., \& {Castets}, A. 2004, \aap, 422, 159, \dodoi{10.1051/0004-6361:20047186}

\bibitem[{{Webber}(1998)}]{Webber1998}
{Webber}, W.~R. 1998, \apj, 506, 329, \dodoi{10.1086/306222}

\end{thebibliography}

\end{document}